\documentstyle[11pt]{article}
\def\hepth#1{ {\tt hep-th/#1}}
\begin{document}
\begin{flushright}
{ ~}\vskip -1in
US-FT/5-99\\
\hepth{9902161}\\
Feb 1999\\
\end{flushright}

\vspace*{20pt}
\bigskip

\centerline{\Large ${\cal N}=2$ SUPERSYMMETRIC YANG-MILLS THEORIES}
\bigskip
\centerline{\Large AND WHITHAM INTEGRABLE HIERARCHIES}
\vskip 0.9truecm
\centerline{\large\sc Jos\'e D. Edelstein ~and~ Javier Mas}

\vskip 0.9truecm

\centerline{\em Departamento de F\'\i sica de Part\'\i culas, 
Universidade de Santiago de Compostela} 

\centerline{\em E-15706 Santiago de Compostela, Spain}

\centerline{\em E-mail: {\tt edels,jamas@fpaxp1.usc.es}}

\vspace{7mm}

\begin{abstract}
We review recent work on the study of ${\cal N}=2$ super Yang-Mills theory with
gauge group $SU(N)$ from the point of view of the Whitham hierarchy,
mainly focusing on three main results: 
(i) We develop a new recursive method to compute the whole instanton
expansion of  the low-energy effective prepotential; 
(ii) We interpret the slow times of the hierarchy as additional couplings
and promote them to spurion superfields that softly break ${\cal N}=2$
supersymmetry down to ${\cal N}=0$ through deformations associated to
higher Casimir operators of the gauge group;
(iii) We show that the Seiberg--Witten-Whitham equations provide a set of 
non-trivial constraints on the form of the strong coupling expansion in
the vicinity of the maximal singularities. 
We use them to check a proposal that we make for the value of the off-diagonal
couplings at those points of the moduli space.
\end{abstract}
\vspace{5mm}

\setcounter{page}{1}

\def\subsub{\par\addtocounter{subsubsection}{1}{\underline{\it
\thesubsubsection}}\hskip0.5cm}
\def\subsubi{{\underline{\it\thesubsubsection}} }

\def\med{{1\ov 2}}
\def\hepth#1{ {\tt hep-th/#1}}

\def\bemat{\left( \begin{array}}
\def\enmat{\end{array} \right)}

\def\beqay{\begin{equation}\begin{array}}
\def\eeqay{\end{array}\end{equation}}

\def\bepf{\bd \item{ {\sl Proof : }} }
\def\enpf{\ed}

\def\bej{\begin{ejer}}
\def\eej{\end{ejer}}

\def\imp{\Rightarrow}
\def\reals{\mathord{\bf R}} %--- reals
\def\comps{\mathord{\bf C}} %--- complex nos.
\def\quats{\mathord{\bf H}} %--- quaternions
\def\integ{\mathord{\bf Z}} %--- integers
\def\rats{\mathord{\bf Q}} %--- rationals
\def\nats{\mathord{\bf N}} %--- naturals
\def\field{\mathord{\bf F}} %--- ground field
\def\Re{{\rm Re}}
\def\Im{{\rm Im}}

\def\be{\begin{equation}}
\def\ee{\end{equation}}
\def\bes{\begin{equation*}}
\def\ees{\end{equation*}}

\def\beqa{\begin{eqnarray}}
\def\beqas{\begin{eqnarray*}}
\def\eeqa{\end{eqnarray}}
\def\eeqas{\end{eqnarray*}}
\def\bea{\begin{eqnarray}}
\def\eea{\end{eqnarray}}

\def\bd{\begin{description}}
\def\ed{\end{description}}

\def\div{{\rm div}}
\def\inv{^{-1}}
\def\cl{\mbox{\tiny (class)}}

\def\etal{{\it et al.}}%--- et al.
\def\ie{{\it i.e.}}%--- i.e.
\def\eg{{\it e.g.}}%--- e.g.
\def\Cf{{\it Cf.\ }}%--- Cf.
\def\cf{{\it cf.\ }}%--- cf.

\def\al{\alpha}
\def\lam{\lambda}
\def\blam{\bar\lambda}
\def\th{\theta}
\def\bth{\bar\theta}
\def\bsigma{\bar\sigma}
\def\bpsi{\bar\psi}

\def\vev{{\em vev} }

\noindent

\def\cross{\, \| \!\!\!\!\!\! =}

\def\H{{\cal H}}
\def\tr{{\rm tr}}
\def\Tr{{\rm Tr}}
\def\F{{\cal F}}
\def\N{{\cal N}}
\def\res{{\rm res}}

\def\d{\partial}
\def\ov{\over}

\def\pder#1#2{{{\partial #1}\over{\partial #2}}}%--- partial derivative
\def\der#1#2{{{d #1}\over {d #2}}}%--- full derivative
\def\ppder#1#2#3{{\partial^2 #1\ov\partial #2\partial #3}}
\def\dpder#1#2{{\partial^2 #1\ov\partial #2 ^2 }}
\def\bemat{\left(\begin{array}}
\def\enmat{\end{array}\right)}
\def\theequation{\arabic{equation}}
\def\thesection{\Roman{section}}

\def\Fpb{\beta\!\cdot\!\!\F'_1}
\def\Fppb{\beta\!\cdot\!\!\F''_1\!\!\cdot\! \beta}
\def\Fpk{\alpha\!\cdot\!\!\F'_k}
\def\Fpu{\alpha\!\cdot\!\!\F'_1}
\def\Fpd{\alpha\!\cdot\!\!\F'_2}
\def\Fppk{\alpha\!\cdot\!\!\F''_k\!\!\cdot\! \alpha}
\def\Fppu{\alpha\!\cdot\!\!\F''_1\!\!\cdot\! \alpha}
\def\Fppd{\alpha\!\cdot\!\!\F''_2\!\!\cdot\! \alpha}
\def\cdotsh{\!\cdot}

\def\comm#1#2{[#1,#2]}

%%%%%%%%%%%%%
%%%%%%%%%%%%%
\section{Introduction}
%%%%%%%%%%%%%
%%%%%%%%%%%%%

The study of non-perturbative phenomena in quantum field theory has experienced 
drastic advances since, in 1994, Seiberg and Witten gave an ansatz for the exact
effective action governing the low-energy excitations of $SU(2)$ ${\cal N} = 2$ 
super Yang-Mills theory \cite{SeiWitt}.
It is given in terms of an auxiliary complex algebraic curve, whose moduli space
is identified with the quantum moduli space of the low-energy theory ${\cal
M}_\Lambda$, and a given meromorphic differential, $dS_{SW}$, that induces
a special geometry  on ${\cal M}_\Lambda$.
Apart form its unquestionable beauty, it proved to contain nontrivial
dynamical information about the non perturbative behaviour of the theory. 
For example, the existence of a confinement mechanism when breaking ${\cal N}=2$ 
to ${\cal N}=1$ by addition of a mass term. 
The solution was soon extended to the case of $SU(N)$ \cite{sun,klemm}.
Still, the  price to pay was the need for ${\cal N}=2$ supersymmetry.   
Soft supersymmetry breaking was shown to preserve the analytic properties of the
solution in such a way that exact results in the ${\cal N}=0$ theory could be
obtained \cite{soft,moresoft,hsu}.

Interestingly enough, it was soon realized that the Seiberg--Witten solution
could be reformulated in terms of certain integrable systems, $dS_{SW}$ being a
solution of their averaged (Whitham) dynamics \cite{gorsketal}. 
For example, the
periodic Toda lattice is the proper integrable system whose averaged dynamics
corresponds to pure ${\cal N} = 2$ super Yang-Mills theory for the whole
ADE series \cite{MWEYNT}. The spectral curve $\Gamma_{g}$ of the
particular integrable system is
identified with the auxiliary Seiberg--Witten algebraic curve.
Its moduli, in spite of being {\em local} invariants, evolve with respect to the
so-called {\em slow times} $T_n$. 
The system of non-linear equations that describe this evolution, that amounts to
adiabatic {\em deformations} of an hyperelliptic curve, was developed by Whitham
\cite{Whitham}.  
Surprisingly, this system turns out to be itself integrable and receives
the generic 
name of {\em Whitham hierarchy} (see \cite{krichever} and references therein). 

The Whitham dynamics can be thought of as a generalization of the Renormalization
Group (RG) flow \cite{gorsketal} (see \cite{Carroll} for a review).
The corresponding RG equations were recently derived by Gorsky, Marshakov, 
Mironov and
Morozov \cite{ITEP}: the second derivatives of the prepotential with respect to
Whitham slow times $T_n$ result to be given in terms of Riemann Theta-functions.
We would like to show, in this talk, that this framework is very fruitful 
both to study many features of the low-energy dynamics of ${\cal N}=2$
super Yang-Mills theory, as well as to implement {\em natural}
generalizations of the Seiberg--Witten solution.

We shall start by giving a telegraphical account of the Seiberg--Witten 
solution to the low-energy dynamics of $SU(N)$ ${\cal N} = 2$ super
Yang-Mills theory and of the
basic ideas involved in the Whitham hierachies. 
We will mainly focus on the fact that they lead naturally to the concept of
a prepotential and establish thereby the concrete link between both formalisms.
Within this framework, we first develop a new recursive method to compute the 
whole instanton expansion of the low-energy effective prepotential. 
Then, we interpret the slow times of the hierarchy as additional couplings 
and promote them to spurion superfields that softly break ${\cal N}=2$
supersymmetry down to ${\cal N}=0$ through deformations associated to
higher Casimir operators of the gauge group.
We discuss in some detail the case of $SU(3)$. 
Finally, we show that the Seiberg--Witten--Whitham equations provide a set of
non-trivial constraints on the form of the strong coupling expansion in
the vicinity of
the maximal singularities. 
We use them to check a proposal that we make for the value of the off-diagonal
couplings at those points of the moduli space.
Most of the work presented in this talk (the first two applications) was
developed by
the authors in collaboration with Marcos Mari\~no \cite{emm}.
The analysis of the strong coupling expansion near the maximal
singularities was done
in \cite{edemas}. Further generalizations, like extensions to other Lie algebras
and/or inclusion of matter, remain interesting problems of research
\cite{Takanos}.

%%%%%%%%%%%%%
%%%%%%%%%%%%%
\section{The Seiberg--Witten solution}
%%%%%%%%%%%%%
%%%%%%%%%%%%%

The classical potential of ${\cal N}=2$ super Yang-Mills theory with a vector
multiplet in the adjoint representation of $SU(N)$ has flat directions.
There is a family of inequivalent ground states that constitutes the {\em
classical
moduli space} ${\cal M}_0$, parametrized by a constant \vev of the scalar field
$\phi$ in the Cartan sub-algebra
\be
<\phi> = \sum_{i=1}^{N-1} a^i H_i = 
\mbox{diag}~(e_1(a^i),\ldots,e_N(a^i)) ~,
\label{vacuum}
\ee
where $e_i(a)=\lambda^i\cdot a$, $\lambda^i$ being the $i$-th fundamental
weight of the Lie algebra $A_{N-1}$.
At a generic point of ${\cal M}_0$, the unbroken gauge symmetry is $U(1)^{N-1}$. 
In fact, for every positive root $\al_+$, a couple of gauge bosons
$W_\mu^{\pm\al_+}$
gets a mass $M_{\alpha_+}(a)=\sqrt 2|\al_+ \cdot a|$ through the Higgs mechanism.
They are BPS states with central charge $Z_{\alpha_+}$,
$Z_\alpha\equiv\alpha\cdot a$
and $a= a^i\alpha_i$ with $\alpha_i$ the simple roots.
Gauge invariant coordinates for ${\cal M}_0$ can be constructed from the
characteristic
polynomial
\be
W_{A_{N-1}}(\lambda,\bar u_k) = \det (\lambda - <\phi>) = \lambda^N - 
\bar u_2(a) \lambda^{N-2} - \bar u_3(a) \lambda^{N-3} - \cdots - \bar u_N(a) ~,
\label{chequation} 
\ee
whose coefficients, $\bar u_k(a)$, are nothing but the Casimir operators.
{\em Each} microscopic theory, characterized by a value of the $\bar u_k$, 
leads to {\em one} effective field theory at low energies.
We may therefore think of the quantum moduli space ${\cal M}_\Lambda$ of
effective 
field theories as being parametrized by $u_k(a)\sim <\Tr\,\phi^k>$.
It will look like a deformation of ${\cal M}_0$, with the {\em quantum
generated scale}
$\Lambda$ as the deformation parameter.
The microscopic relations, thus, should be recovered for $\Lambda\to 0$.

The low-energy effective action can be entirely written in terms of a
single {\em holomorphic} function of the Cartan variables, $\F(a^i)$,
known as the effective prepotential. 
By means of symmetry considerations \cite{SeiWitt,sun,klemm}, the most general
expression for $\F(a^i)$ is
\be
\F = {1\ov 2N}\tau_0 \sum_{\alpha_+} Z_{\alpha_+}^2
+ {i\ov 4 \pi} \sum_{\alpha_+} Z_{\alpha_+}^2 \log \, {Z_{\alpha_+}^2\ov
\Lambda^2}
+{1\ov 2 \pi i}\sum_{k=1}^\infty \F_k(a)
\Lambda^{2Nk} ~,
\label{effprep}
\ee
where $\tau_0$ is the bare coupling constant and $\F_k(a)$ are Weyl invariant
combinations of $a^i$, whereas $k$ is the instanton number.
The full prepotential is homogeneous of degree two in $a^i$ and $\Lambda$.
What remains to be computed are the instanton corrections, $\F_k(a)$.
It is their exact determination the whole point of the Seiberg--Witten solution.
In the semiclassical limit, the first few terms can be computed
explicitely in the
microscopic theory \cite{hunter}, and their agreement with the output of the
Seiberg--Witten solution provides a non-trivial consistecy check of it.

Since the effective prepotential (\ref{effprep}) is holomorphic, the imaginary
part of its second order derivatives with respect to the Cartan variables,
${\rm Im}\tau_{ij}$, are harmonic functions. Therefore, they cannot have a global
minimum.   However, ${\rm Im}\tau_{ij}$ enters the low-energy Lagrangian as an
effective coupling constant: it {\em must} be positive definite. So,
$\tau_{ij}$ cannot
be globally defined. 
These conditions are automatically fulfilled it $\tau_{ij}$ is the
period matrix  of some Riemann surface.
This observation led Seiberg and Witten to the following ansatz:

\noindent 
{\bf First:} Over each point on ${\cal M}_\Lambda$ labelled by $u_k$, consider a 
certain hyperelliptic curve. 
For $SU(N)$ the  relevant curve $\Gamma_g$ is \cite{sun}
\be
y^2 = P(\lambda, u_k)^2 - 4\Lambda^{2N}
\label{hyper}
\ee
with $P = W_{A_{N-1}}$. 
The moduli space of the family of Riemann surfaces of genus $g=N-1$ written above
is to be identified with ${\cal M}_\Lambda$. 
In the classical limit $\Lambda\to 0$, the rational function
$y = W_{\,A_{N-1}}(\lambda, u_k)$ has roots $e_i(a)$. 
At the quantum level, as $y^2$ factors into the product $y^2= y_+ y_-$ with
$y_\pm=(P(\lambda,u_k)\pm 2\Lambda^N)$, the points $e_i$ split into two sets 
of roots of $y_\pm$,
\be
e_i(\bar u_k) \to\ e_i^\pm(u_k,\Lambda) \equiv e_i(u_{k<N},u_N\pm 2\Lambda^N) ~,
\label{split}
\ee
that become the $2N$ branch points of the Riemann surface.

\noindent
{\bf Second:} At the point $u_k$ on ${\cal M}_\Lambda$, the (quantum)
relations between 
$a^i, a_{D\,j}$ and $u_k$ is given by the period integrals
\be
a^i(u) = \oint_{A^i} dS_{SW}(u) ~~~~~~~~~~
a_{D\,j}(u) = \oint_{B_j} dS_{SW}(u) ~.
\label{theper}
\ee
with $dS_{SW}$ a meromorphic differential given by
\be
dS_{SW} =  {\lambda P'(\lambda,u_k)\ov
\sqrt{P^2(\lambda,u_k) - 4\Lambda^{2N}}} ~d\lambda ~,
\label{oneform}
\ee
and $A^i$ and $B_j$ constitute a symplectic basis of homology cycles of the
hyperelliptic curve, with the canonical intersections $A^i\cap A^j = B_i\cap B_j 
= 0$ and $A^i\cap B_j = \delta^i{_j}$, $i,j=1,\dots,N-1$. 
The prepotential $\F(a)$ is implicitely defined by the equation
\be
a_{D\,i} = \pder{\F(a)}{a^i}~.
\ee
The exact determination of $\F(a)$ involves, in general, the integration of
functions $a_{D\,i} (a)$ for which there is not a closed form available.
Therefore, $\F(a)$ will only be calculable in a series expansion.

\noindent
{\bf Third:} The BPS spectrum is obtained by integrating the Seiberg--Witten
differential $dS_{SW}$ along all nontrivial cycles of the Riemann surface, 
$\nu(n^e,n_m) = n^e\cdot A + n_m\cdot B$. 
In fact, this is immediate from the previous point and the fact that the
central charge
of a state with $n^e{_i}$ units of electric charge and $n_m{^i}$ units of
magnetic
charge with respect to the $i$-th $U(1)$ unbroken  subgroup can be written as
\be
Z(n^e,n_m) = n^e \cdot a + n_m \cdot a_D ~.
\label{central}
\ee
Appart from the mass, what remains invariant is the {\em intersection
number} of two BPS states, given by the intersection product of their respective
cycles $\nu(n^e,n_m)$ and $\nu'({n'}^e,{n'}_m)$
\be
\nu\cap\nu' = n^e \cdot {n'}_m - {n'}^e \cdot n_m \in\integ ~.
\label{zwa}
\ee
Notice that this is nothing but the Dirac-Schwinger-Zwanzinger
quantization condition
\cite{DSZ}. 
Two dyons are mutually local if they have zero intersection $\nu\cap\nu'=0$.

Changes in the symplectic basis of homology cycles are performed by means of a
symplectic matrix $\Gamma\in Sp(2r,\reals)$. 
Accordingly, ${\bf a}= (a^i,a_{D\,j})$ transforms as a vector and
$\tau_{ij}$ as a modular form.
Since the central charge, $Z_{\bf n} = {\bf n}^t\cdot {\bf a}$ with ${\bf n}=
(n^e_i,n_m^j)$, is an observable, the invariance
of the {\em non-perturbative} BPS spectrum breaks the continuous duality group
$Sp(2r,\reals)$ down to the discrete subgroup $Sp(2r,\integ)$. 

\noindent
{\bf Fourth:} There are singularities in ${\cal M}_\Lambda$, encoded in
the quantum
discriminant $\Delta_\Lambda$,
\be
\Delta_\Lambda(u_k,\Lambda) = \prod_{i<j}(e_i^+-e_j^+)^2(e_i^--e_j^-)^2
= {\rm c} ~\Lambda^{2N^2} \Delta_+ \Delta_- ~,
\label{quantdisc}
\ee
at whose zero locus, $\Sigma_\Lambda$, two branch points $e_i^\pm$,
$e_j^\pm$ collide.
What is the same, a certain homology cycle on the Riemann surface shrinks
to zero at $\Sigma_\Lambda$, signaling the appearance of an extra massless
state which is generically a dyon. 
These singularities lie on curves that intersect at points where many BPS
states become simultaneously massless. In particular, at the so-called
$N-1$ points, exactly $N-1$
mutually local monopoles become massless. This is the maximal number of
mutually local
simultaneously massless BPS states. The physics of these points was first
investigated
in Ref.\cite{ds}. 
They remain the vacua of the ${\cal N}=1$ theory upon perturbation of
the ${\cal N}=2$ theory by a mass term.
For $SU(N)$, $N>2$, there are regions in ${\cal M}_\Lambda$ where {\it mutually
non-local} dyons become simultaneously massless, and the corresponding effective
low-energy dynamics seems to be given by a superconformal field theory \cite{ad}.

%%%%%%%%%%%%%
%%%%%%%%%%%%%
\section{The universal Whitham hierarchy}
%%%%%%%%%%%%%
%%%%%%%%%%%%%

The name {\em Whitham hierarchy} stands for a wide class of integrable systems
of differential equations that describe modulations of solutions of soliton
equations \cite{Whitham,FFM,oldKri}.
Following Krichever \cite{krichever}, we define the moduli space of the Whitham
hierarchy by 
\be
\hat {\cal M}_{g,p} \equiv
\{\Gamma_g,P_a,\xi_a(P),~a=1,...,p\}
\label{moduliwh}
\ee 
containing the following set of algebraic--geometrical data:

\noindent $\bullet$ $\Gamma_g$ denotes a smooth algebraic curve of genus $g$. 

\noindent $\bullet$ $P_a$ is a set of $p$ points (punctures) on $\Gamma_g$
in generic
positions (we will consider, for simplicity, $p=1$).

\noindent $\bullet$ $\xi_a $ are local coordinates in the neighbourhood of
the $p$ points, \ie\ $\xi_a(P_a)= 0$.

From the general theory of meromorphic differentials over Riemann surfaces
we know
that there are three basic types of Abelian differentials:

\noindent i. {\sl Holomorphic differentials,} $dw_i$.~ 
In any open set $U\in\Gamma_g$, with complex coordinate $\xi$, they are of
the form 
$dw = f(\xi) d\xi$ with $f$ an holomorphic function.
The vector space of holomorphic differentials on a genus $g$ Riemann surface has 
complex dimension $g$.
If the curve is hyperelliptic (\ref{hyper}), a canonical basis $\{dw_j\}$
of this 
vector space is defined through
\be
\oint_{A^i} dw_j = \delta^i{_j} ~~~~~~~~~~~~~~~~~
\oint_{B_i} dw_j = \tau_{ij} ~,
\label{ciclos}
\ee
where $\tau_{ij}$ is the period matrix of the complex curve.

\noindent ii. {\sl Meromorphic differentials of the second kind,}
$d\Omega_{P,n}$.~ 
They have a single pole of order $n+1$ at point $P\in\Gamma$, and zero residue. 
In local coordinates $\xi$, we shall adopt the normalization
\be
d\Omega_{P,n} = (\xi^{-n-1} + O(1)) ~d\xi ~.
\label{merseco}
\ee 
This fixes $d\Omega_{P,n}$ up to an arbitrary combination of holomorphic
differentials.
There are several ways to fix this normalization.
In the context of integrable theories, the standard way to do it is to
require that
$d\Omega_{P,n}$ has vanishing $A^i$-periods
\be
\oint_{A^i} d\Omega_{P,n} = 0 ~.
\label{mersecod}
\ee

\noindent iii. {\sl Meromorphic differentials of the third kind,}
$d\Omega_{P,0}$.~
They have first order poles at $P$ and $P_0$ (a reference point) with
opposite residues
taking values $+1$ and $-1$ respectively.
In local coordinates $\xi\,(\xi_{0})$ about $P\,(P_0)$,
\be
d\Omega_{P,0} = (\xi^{-1} +{\cal O}(1)) d\xi =  
-(\xi_0^{-1} +{\cal O}(1)) d\xi_0 ~.
\label{merter}
\ee
The regular part is normalized by demanding that $d\Omega_{P,0}$ has vanishing 
$A^i$-periods.
The appearance of simple poles in the Seiberg--Witten solution is related to the
inclusion of matter hypermultiplets in the fundamental representation 
\cite{SeiWitt}. 
We will only consider in this talk the case of {\em pure} $SU(N)$, ${\cal
N}=2$ super Yang-Mills theory.
Thus, we are going to rule out the meromorphic differentials of the third 
kind from our discussion.

The standard Whitham equations take the following form \cite{oldKri}
\be
\pder{d\Omega_{n}}{T^m} = \pder{d\Omega_{m}}{T^n}
\label{Whitone}
\ee
where $d\Omega_n$ is short for $d\Omega_{P,n}$, and $T^n$ are a set of
{\em slow times}
the 1-forms may depend upon.
According to our previous remark, $n$ will be considered greater or
equal than one, unless the contrary is stated.
The Whitham hierarchy can be enhanced to incorporate also holomorphic
differentials
$dw_i$, with associated parameters $\alpha^i$, such that
\be
\pder{dw_i}{\alpha^j} =\pder{dw_j}{\alpha^i}  ~~~~~~~~~~~~
\pder{dw_i}{T^n} = \pder{d\Omega_{n}}{\alpha^i} ~~~~~~~~~~~~
\pder{d\Omega_{n}}{T^m} = \pder{d\Omega_{m}}{T^n} ~.
\label{Whithext}
\ee
Equations (\ref{Whithext}) are nothing but the integrability conditions
implying the
existence of a {\em generating differential} $dS$ satisfying
\be
\pder{dS}{\alpha^i} = dw_i ~~~~~~~~~~~~~~~~~~~~~ \pder{dS}{T^n} = d\Omega_{n} ~.
\label{dese}
\ee
The Whitham equations hide a certain holomorphic function named {\em
prepotential} $\F(\alpha^i, T^n)$, that can be defined implicitely through
the following set of equations
\be
\pder{\F}{\alpha^j} = \oint_{B_j} dS ~~~~~~~~~~~~~~~~~~~~~~~~
\pder{\F}{T^n} = {1\ov 2\pi i n}\oint_P \xi^{-n} dS ~.
\label{theo1}
\ee
The local behaviour of the generating differential near the puncture $P$ is then
\be
dS \sim \left\{ \sum_{n\geq 1} T^{n} \xi^{-n-1} +
2\pi i \sum_{n\geq 1} \, n\pder{\F}{T^{n}} \xi^{n-1} \right\} d\xi ~.
\label{local}
\ee
An interesting {\em class of solutions}, and certainly that which is relevant in
connection with ${\cal N}=2$ super Yang-Mills theories, is given by those
prepotentials 
that are homogeneous of degree two:
\be
\sum_{i=1}^{N-1}\alpha^i\pder{\F}{\alpha^i} + \sum_{n\geq 1} 
T^{n}\pder{\F}{T^{n}} = 2\F ~.
\label{homodos}
\ee
The generating differential $dS$ for homogeneous solutions admits the following 
form \cite{MWEYNT,krichever}:
\be
dS = \sum_{i=1}^{N-1} \alpha^i dw_i + \sum_{n\geq 1} T^{n} d\Omega_{n} ~,
\label{desedos}
\ee
and, after (\ref{ciclos})--(\ref{mersecod}), the parameters $\alpha^i$,
and $T^{n}$ can be recovered from $dS$ as follows:
\be
\alpha^i =  \oint_{A^i}dS ~~~~~~~~~~~~~~~~~~~~~
T^{n}  = \res_{P}\,\xi^{n} dS ~.
\label{lospara}
\ee
Inserting (\ref{dese}) and (\ref{lospara}) into (\ref{homodos}), a formal
expression for $\F$ in terms of $dS$ can be obtained \cite{itomoro},
\be
\F = \med\sum_{i=1}^{N-1}\oint_{A^i}dS\oint_{B_i}dS + {1\ov 4\pi i}
\sum_{n\geq 1}{1\ov n}\oint_{P}\xi^{n}dS\oint_{P}\xi^{-n} dS ~.
\label{explicit}
\ee
Following \cite{itomoro}, let us consider the decomposition of $dS$ in a
different basis of Abelian differentials,
\be
dS= \sum_{n\geq 1} T^{n} d\hat\Omega_{n} ~,
\label{deseuno}
\ee
where $d\hat\Omega_{n}$ are meromorphic differentials of the second kind
(with the same local behaviour than $d\Omega_{n}$), whose regular part is
fixed by the condition:
\be
\pder{d\hat\Omega_{n}}{\hbox{moduli}} = \hbox{holomorphic} ~.
\label{defholo}
\ee
Notice that we have not added explicitely holomorphic differentials in $dS$:
they are somehow hidden inside the differentials $d\hat\Omega_{n}$.
In more concrete terms, the definition of the $\alpha^i$ parameters as given in
(\ref{lospara}), now forces them to depend on $T^n$ and $u_k$.
Conversely, provided we impose that $d\alpha^i/dT^n= 0$, an implicit set of
homogeneous functions $u_k(T^n,\alpha^i)$ of degree zero is obtained, and
they solve the Whitham equations:
\be
\pder{u_k}{T^n} = -\left(\pder{\alpha^i}{u_k}\right)\inv\pder{\alpha^i}{T^n} ~.
\label{whieq}
\ee
Finally, from (\ref{lospara}) and (\ref{deseuno}), it is clear that
\be
d\hat\Omega_m = d\Omega_m + \sum_{i=1}^{N-1} \pder{\alpha^i}{T^m} dw_i ~.
\label{relation}
\ee

%%%%%%%%%%%%%
%%%%%%%%%%%%%
\section{The Seiberg--Witten--Whitham Formalism}
%%%%%%%%%%%%%
%%%%%%%%%%%%%

The next task is to look for an embedding of the Seiberg--Witten ansatz within
the Whitham hierarchy.
We will follow very closely, to this end, the approach of Gorsky,
Marshakov, Mironov and Morozov \cite{ITEP}. 
As already noticed in Ref.\cite{gorsketal}, the curve
(\ref{hyper}) is the hyperelliptic representation for the spectral curve
of the {\em periodic Toda chain} of length $N$. It can be written in terms
of a complex parameter
$w$ as follows (we set $\Lambda=1$ for convenience)
\be
P =  \left( w + {1\ov w}\right) ~~~~~~~~~~
y =  \left( w - {1\ov w}\right) ~.
\label{Py}
\ee
This defines a natural coordinate in the vicinity of the two points at infinity
$\infty_\pm\sim(\pm y=\infty,\lambda=\infty)$.
In fact, from Eq.(\ref{Py}), $w$ can be written as a meromorphic function
\be
w = \med (P+y) ~,
\label{functionw}
\ee
that near $\infty_\pm$ goes as $w\sim\lambda^{\pm N}$.
Then, $\xi_\pm = w^{\mp 1/N}$ are local coordinates at the punctures
$\infty_\pm$, which are the points where the relevant meromorphic
differential of the Seiberg--Witten
solution, $dS_{SW}$, has its (second order) poles.
The times associated to each puncture will be denoted with positive and negative
subindices, \ie\ $T_{\infty_\pm,n} = T_{\pm n}$. 
Also, it is convenient to slightly change the normalization of our second-kind
differentials to be
$d\Omega_{\pm n} \sim {N\ov n}\, dw^{\pm n/N}$.

The Seiberg--Witten differential, $dS_{SW}$, belongs to the 
{\em class of solutions} 
of the Whitham hierarchy that fulfill
\be
\pder{dS}{\hbox{moduli}} = \hbox{holomorphic} ~.
\label{swhol}
\ee
In fact, since its $A^i$-periods are $a^i$, one is tempted to identify
$dS_{SW}$ as
the generating form of the Whitham hierarchy at $\alpha^i = a^i$ and
$T_{\pm1} = 1,~
T_{|n|>1}=0$,
\be
dS_{SW} = a^i dw_i + d\Omega_{\infty_+,1} + d\Omega_{\infty_-,1} ~.
\label{desw}
\ee
Varying Eq.(\ref{Py}) for a given curve, \ie\ for fixed $u_k$ and $\Lambda$, the
Seiberg--Witten differential can be written as
\be
dS_{SW} = {\lambda P'\ov y}d\lambda = \lambda{dw\ov w} ~,
\label{deswbis}
\ee
this making extremely easy to verify the defining property of $dS_{SW}$,
\be
\left.\pder{dS_{SW}}{u_k}\right\vert_{w=const.} =  {\lambda^{N-k}\ov P'}{dw\ov w}
= {\lambda^{N-k}d\lambda\ov y} = dv_k~~,~~~~~~~~k=2,3,...,N ~.
\label{derdeese}
\ee
If we hold $\lambda$ fixed --instead of $\omega$--, there is an additional total
derivative in the previous expression.
Notice that, from the point of view of the Whitham equations, it matters which 
coordinates are held  fixed as long as there are residues to be computed.
It will always be understood that derivatives w.r.t. the moduli are taken at 
constant $w$.

{\bf Lemma} \cite{ITEP}: The meromorphic differentials $d\hat\Omega_n$ have
the form
\be
d\hat\Omega_n = R_n{dw\ov w} = P_+^{n/N} {dw\ov w} ~,
\label{lema}
\ee
where the projection $(\sum_{k=-\infty}^\infty c_k\lambda^k)_+ =
\sum_{k=0}^\infty c_k\lambda^k$.
Notice that, from their defining equation (\ref{defholo}), the
$d\hat\Omega_n$ have 
poles at both punctures, \ie\ there are not $d\hat\Omega_{\pm n}$ but
$d\hat\Omega_n = d\hat\Omega_{n} + d\hat\Omega_{-n}$.

At this point, as shown in Ref.\cite{ITEP}, the first and second order
derivatives of the prepotential can be computed from Eq.(\ref{theo1}),
with the following result
\be
\pder{\F}{T_n} = {\beta\ov 2\pi i n} \sum_{m} m T_m\H_{m+1,n+1} ~~~~~~~~~~~~~~~
\ppder{\F}{\alpha^i}{T^n} = {\beta\ov 2\pi i n} \pder{\H_{n+1}}{a^i} ~, 
\label{completeness}
\ee
\be
\ppder{\F}{T_m}{T_n} = -{\beta\ov 2\pi i} \left(
\H_{m+1,n+1} + {\beta\ov mn}\pder{\H_{m+1}}{a^i}\pder{\H_{n+1}}{a^j}
{1\ov i\pi}\d_{\tau_{ij}}\log \Theta_E(0|\tau)\right) ~,
\label{rgequ}
\ee
where $\beta = 2N$ is the coefficient of the beta function, whereas
$\H_{m+1,n+1}$ and $\H_{n+1}$ are homogeneous combinations of the Casimirs
constructed as follows:
\be
\H_{m+1,n+1} = {N\ov mn} \mbox{res}_\infty
\left(P^{m/N}(\lambda)dP_+^{n/N}(\lambda)\right) = \H_{n+1,m+1} 
~~~~~~~~~~~~~~~~~~~~~ \H_{m+1} \equiv \H_{m+1,2} ~.
\label{hamimn}
\ee
The important ingredient is the Riemann's Theta function $\Theta_E(\vec
z|\tau)$, $E$
being the following even and half-integer characteristic \cite{emm}:
\be
\vec \alpha=(0, \dots, 0) ~~~~~~~~~~~~~~~~~~~~~~
\vec \beta = (1/2, \dots, 1/2) ~.
\label{carac}
\ee
As they stand, however, the expressions given in
(\ref{completeness})--(\ref{rgequ}) 
are not yet suitable for application to the Seiberg--Witten solution.
It is still necessary to define the rescaled times $\hat T_n = T_1^{-n}T_n$ and
moduli $\hat u_k = T_1^k u_k$ (correspondingly, $\hat\H_{m+1,n+1} =
T_1^{m+n}\H_{m+1,n+1}$) after which, the prepotential of the Seiberg--Witten 
solution is obtained by identifying $T_1$ with $\Lambda$ in the
submanifold $\hat T_{n>1} = 0$, provided that the moduli space be
parametrized by the $\hat u_k$
(notice that $\hat a^i \equiv \alpha^i(u_k,T_1,\hat T_{n>1} = 0) = T_1a^i(u_k,1)
= a^i(\hat u_k,T_1)$) \cite{emm}. The restriction to the submanifold
$\hat T_{n>1} =0$, yields formulae which are suited for the
Seiberg--Witten solution.
In particular, from (\ref{completeness}) and (\ref{rgequ}) we obtain
\be
\pder{\F}{\log\Lambda\,} = {\beta\ov 2\pi i } \hat\H_{2} ~~~~~~~~~~~~~~~~~~~~~~
\pder{\F}{\hat T_n} = {\beta\ov 2\pi i n}  \hat\H_{n+1} ~,
\label{firstorder}
\ee
\be
\ppder{\F}{\hat a^i}{\log\Lambda} = {\beta\ov 2\pi i}\pder{\hat\H_{2}}{\hat
a^i} ~~~~~~~~~~~~~~~~~~~ \ppder{\F}{\hat a^i}{\hat T_n} = {\beta\ov 2\pi i
n}\pder{\hat\H_{n+1}}{\hat a^i} ~, 
\label{mixed}
\ee
\be
\dpder{\F}{(\log\Lambda)} = -{\beta^2\ov 2\pi i}
\pder{\hat\H_{2}}{\hat a^i}\pder{\hat\H_{2}}{\hat a^j}{1\ov i\pi}
\d_{\tau_{ij}}\log\Theta_E(0|\tau) ~, 
\label{lasecudef} 
\ee
\be
\ppder{\F}{\log\Lambda\,}{\hat T_n} = -{\beta^2\ov 2\pi i n}
\pder{\hat\H_{2}}{\hat a^i}\pder{\hat\H_{n+1}}{\hat a^j}{1\ov i\pi}
\d_{\tau_{ij}}\log\Theta_E(0|\tau) ~, 
\label{lambt}
\ee
\be
\ppder{\F}{\hat T_m}{\hat T_n} = -{\beta\ov 2\pi i} \left(
\hat\H_{m+1,n+1}+{\beta\ov mn}
\pder{\hat\H_{m+1}}{\hat a^i}\pder{\hat\H_{n+1}}{\hat a^j}{1\ov i\pi}
\d_{\tau_{ij}}\log\Theta_E(0|\tau) \right) ~, 
\label{secorder}
\ee
with $m,n\geq 2$. 

The first equation in (\ref{firstorder}) is precisely the RG equation
derived in Ref.\cite{matone}.
Combining the second equation in (\ref{firstorder}) and (\ref{lambt}), it is easy
to obtain an interesting relation between Casimir operators \cite{ITEP}:
\be
\pder{\hat\H_{m}}{\log\Lambda\,} = -\beta\pder{\hat \H_2}{\hat
a^i}\pder{\hat\H_{m}}{\hat
a^j} {1\ov i\pi}\d_{\tau_{ij}} \log \Theta_E(0|\tau) ~.
\label{qucas}
\ee
Hereafter, we will always work with the scaled coordinates and hats will be 
omitted everywhere.

%%%%%%%%%%%%%
%%%%%%%%%%%%%
\section{Instanton Corrections}
%%%%%%%%%%%%%
%%%%%%%%%%%%%

Instanton calculus provides one of the few non-perturbative links between the
Seiberg--Witten solution and the microscopic non-abelian field theory that it is
supposed to describe effectively at low energies.
From the microscopic theory point of view, instanton contributions to the
asymptotic semiclassical expansion of the effective prepotential have been
computed, and a remarkable agreement with the Seiberg--Witten solution has
been found \cite{hunter}.
We shall see in this section that the connection of $SU(N)$ ${\cal N} = 2$ super
Yang--Mills theory with Toda--Whitham hierarchies embodies in a natural
way a recursive procedure to compute the instanton expansion of the
effective prepotential up to arbitrary order.

To begin with, let us fix our conventions.
We choose the basis $H_k = E_{k,k}- E_{k+1,k+1}$  for the Cartan subalgebra
and $E_{k,j},k\neq j$ for the raising and lowering operators.
Let $\{ \alpha_i\}_{i=1,...,N-1}$ stand for the simple roots of $SU(N)$ and
$(\alpha,\beta)$ denote the usual inner product constructed with the
Cartan-Killing form. The dot product $\alpha\cdotsh\beta \equiv
2(\alpha,\beta)/(\beta,\beta)= (\alpha,\beta^\vee)$.
We have that $\alpha_i\cdotsh\alpha_j = C_{ij}$, with $C_{ij}$ the Cartan
matrix,
while $\lambda^i\cdotsh\alpha_j = \delta^i{_j}$ define the fundamental weights.
In particular this means that $\alpha_i = \sum_j C_{ij}\lambda^j$.
The simple roots generate the root lattice $\Delta = \{\alpha = n^i
\alpha_i| n^i\in {\bf Z}\}$.

The instanton expansion of the prepotential was given in eq.(\ref{effprep}).
We then have, for the LHS of (\ref{lasecudef}),
\be
\dpder{\F}{(\log \Lambda)} = {1\ov 2\pi i}
\sum_{k=1}^\infty (2N k)^2\F_k(Z) \Lambda^{2Nk} ~.
\label{makdj}
\ee
The derivative of the quadratic Casimir also has an  expansion that
can be obtained from the RG equation (first equation in
(\ref{firstorder})) and the expansion of the prepotential
\be
\pder{\H_2}{a^i} = {2\pi i\ov \beta} \ppder{\F}{a^i}{\log\Lambda\,}
= C_{ij} a^j +  \sum_{k=1}^\infty  k \F_{k,i} ~ \Lambda^{2Nk} \equiv
\sum_{k=0}^\infty H_i^{(k)} \Lambda^{2Nk} ~,
\label{expanh}
\ee
where $\F_{k,i}= \d\F_k/\d a^i$. 
The term involving the couplings that appear in the Theta function $\Theta_E$ is
\be
i\pi \, n^i\tau_{ij} n^j = \sum_{\alpha_+} \log
\left({Z_\alpha\ov \Lambda}\right)^{-(\alpha\cdot\alpha_+)^2} + \med
\sum_{k=1}^\infty~
(\Fppk) ~\Lambda^{2Nk} ~,
\label{tauroot}
\ee
where $\alpha= n^i\alpha_i$ and
\be 
\Fppk \equiv \sum_{i,j}n^i\ppder{\F_k}{a^i}{a^j}n^j =
\sum_{\beta, \gamma\in \Delta}(\alpha\cdotsh\beta)
\ppder{\F_k}{Z_{\beta}}{Z_{\gamma}}(\gamma\cdotsh\alpha) ~.
\label{efeseg}
\ee
For convenience, we have adjusted the bare coupling to $2\pi i\tau_0 = 3N$.
We may shift $\tau_0$ to any value by an appropriate rescaling of $\Lambda$.
This will be reflected in the normalization of the $\F_k$. 

Inserting (\ref{tauroot}) in the Theta function, we obtain
\beqa
\Theta_E(0|\tau) &=& 
\sum_{r=0}^\infty \sum_{\alpha\in\Delta_r} (-1)^{\rho\cdot\alpha}
\prod_{\alpha_+} Z_{\alpha_+}^{-(\alpha\cdot\alpha_+)^2}
\prod_{k=1}^\infty\left(\sum_{m=0}^\infty {1\ov 2^m
m!}\left(\Fppk\right)^m\, \Lambda^{2Nkm}\right) \Lambda^{2Nr}
\cr &\equiv& \sum_{p=0}^\infty \Theta^{(p)} \Lambda^{2Np} ~, 
~~~~~~~~~~~~~~~~~~~~~~~~~~ \rho = \sum_{i=1}^{N-1}\lambda^i ~.
\label{expantheta}
\eeqa
In the previous expression, $\Delta_r\subset\Delta$ is a subset of the
root lattice
composed of those lattice vectors $\alpha$ that fulfill the constraint
$\sum_{\alpha_+}(\alpha\cdot\alpha_+)^2 = 2Nr$.
In particular $\Delta_1$ is the root system, {\it i.e.} the simple roots
together with their Weyl reflections.
On the other hand $\Delta_r$, for $r>1$, will be in general a union of
Weyl orbits,
since Weyl reflections are easily seen to be an automorphisms of $\Delta_r$.
Therefore, $\Theta^{(p)}$ is Weyl invariant by construction. 
In the logarithmic derivative, $\Theta_E$ appears in the denominator, so
we need the
expansion of the inverse of the Theta function (see Ref.\cite{emm} for details):
\be
\Theta(0|\tau)^{-1} = \sum_{l=0}^{\infty} \Xi^{(l)}(\Theta) \,
\Lambda^{2Nl} ~.
\label{tauinv}
\ee
Finally, the derivative of the Theta function with respect to the period
matrix is given by
\beqa 
{1\ov i\pi}\d_{\tau_{ij}}\Theta_E(0,\tau) & = & \sum_{r=1}^\infty 
\sum_{\alpha\in\Delta_r} (-1)^{\rho\cdot\alpha} 
(\alpha\cdotsh\lambda^i)(\alpha\cdotsh\lambda^j)
\prod_{\alpha_+} Z_{\alpha_+}^{-(\alpha\cdot\alpha_+)^2}
\prod_{k=1}^\infty\exp\left({\med (\Fppk) \Lambda^{2Nk}}\right)
\Lambda^{2Nr} \cr
& \equiv & \sum_{p=1}^\infty \Theta_{ij}^{(p)} \Lambda^{2Np} ~.
\label{expanthij}
\eeqa
Collecting all the pieces and inserting them back into (\ref{lasecudef}),
we find for $\F_k(Z)$ the following expression:
\be
\F_k(Z) = - k^{-2}
\sum_{p, q, l=0}^{p+q+l = k-1}\sum_{ij} H_i^{(p)} H_j^{(q)}
\Theta_{ij}^{(k-p-q-l)} \Xi^{(l)} ~,
\label{elresul}
\ee
in terms of the previously defined coefficients.
If we look at the coefficients in the r.h.s. of Eq.(\ref{elresul}), it is
easy to see
that the expressions they involve depend on $\F_1,\F_2,...$ up to $\F_{k-1}$.
In fact, although both $H^{(p)}$ and $\Theta^{(p)}$ depend on
$\F_1,....\F_p$, the
indices within parenthesis reach at most the value $k-1$ as
$\Theta_{ij}^{(0)}=0$.
Moreover $\Theta_{ij}^{(k)}$ depends on $\F_1,...,\F_{k-1}$ since the
vector $\alpha=0$
is missing from the lattice sum.
This fact implies the possibility to build up a recursive procedure to
{\em compute all the instanton coefficients} by starting just from the
perturbative contribution to
$\F(a)$ in (\ref{effprep}).
The first few instanton contributions, for example, are simply \cite{emm}:
\be
\F_1 = - \sum_{\alpha\in\Delta_1}(-1)^{\rho\cdot\alpha} Z_\alpha^2 
\prod_{\alpha^+} Z_{\alpha^+}^{-(\alpha\cdot\alpha^+)^2} ~,
\label{efe1}
\ee
\beqa
\F_2 & = & - {1\ov 4}\left(
\sum_{\alpha\in\Delta_1}(-1)^{\rho\cdot\alpha}
\prod_{\alpha_+} Z_{\alpha_+}^{-(\alpha\cdot\alpha_+)^2}
\left[ \F_1 + 2 (\Fpu ) Z_\alpha + \med (\Fppu ) Z_\alpha^2 \right] \right. 
\nonumber \\ 
& & \left. + \sum_{\beta\in\Delta_2}Z_\beta^2~ 
(-1)^{\rho\cdot\beta}\prod_{\alpha^+}
Z_{\alpha^+}^{-(\beta\cdot\alpha^+)^2} \right) ~,
\label{efe2}
\eeqa
\beqa
\F_3 & = & - {1\ov 9}\left(
\sum_{\alpha\in\Delta_1}(-1)^{\rho\cdot\alpha} \prod_{\alpha^+}
Z_{\alpha^+}^{-(\alpha\cdot\alpha^+)^2} \left[\rule{0mm}{5mm} 4\F_2 + 4 (\Fpd) 
Z_\alpha + (\Fpu)^2 + \med (\Fppu)\left( \F_1 + 2 (\Fpu) Z_\alpha 
\right)\right.\right.
\nonumber \\ & & \left.\left. + {1\ov 8} (\Fppu)^2 Z_\alpha^2 + \med (\Fppd) \,
Z_\alpha^2 \right] + \sum_{\beta\in\Delta_2}(-1)^{\rho\cdot\beta}
\prod_{\alpha_+} Z_{\alpha_+}^{-(\beta\cdot\alpha_+)^2} \left[ \F_1 + 2 (\Fpb) 
Z_\beta  \right. \right. \nonumber \\ 
& & \left. \left. + \med (\Fppb ) Z_\beta^2 \right] +
\sum_{\gamma\in\Delta_3}(-1)^{\rho\cdot\gamma}
\prod_{\alpha^+} Z_{\alpha^+}^{-(\gamma\cdot\alpha^+)^2}\,
Z_\gamma^2\!\rule{0mm}{7mm}\right) ~,  
\eeqa
etc.
The above expressions make patent the recursive character of the procedure.

%%%%%%%%%%%%%
%%%%%%%%%%%%%
\section{Soft SUSY Breaking with Higher Casimir Operators}
%%%%%%%%%%%%%
%%%%%%%%%%%%%

Sofly--broken supersymmetric models offer the best phenomenological candidates 
to solve the hierarchy problem in grand--unified theories.
The {\em spurion formalism} \cite{spurion} provides a tool to generate soft
supersymmetry breaking in a neat and controlled manner. To illustrate
the method, start from a supersymmetric lagrangian, $L(\Phi_0,\Phi_1,...)$
with some set of chiral superfields, and single out  a particular one, say
$\Phi_0$. 
If you let this superfield acquire a constant \vev along a given direction in 
superspace like, for example, $<\Phi_0>= c_0+ \theta^2 F_0$, it will induce soft
breaking terms, and a vacuum energy of order $|F_0|^2$.
Turning the argument around, you could {\em promote} any parameter in your
lagrangian to a chiral superfield, and then {\em freeze} it along a
supersymmetry breaking direction in superspace giving a \vev to its
highest component.

From embedding the Seiberg--Witten solution within the Toda--Whitham
framework, we 
have obtained the analytic depence of the prepotential on some new
parameters $T_n$. 
In this section, we will interpret these slow times as parameters of a
non-supersymmetric family of theories, by promoting them to spurion superfields. 
In Refs.\cite{soft,moresoft,luisIyII} this program was initiated with the scale
parameter $\Lambda$ and the mases of additional hypermultiplets, $m_i$, as
the only
sources for spurions.
The slow times, as Eq.(\ref{firstorder}) shows, are dual to the
$\H_{m+1}$, which are
homogeneous combinations of the Casimir operators of the group. 
This means that we will be able to parametrize soft supersymmetry breaking
terms induced
by all the Casimirs of the group, and not just the quadratic one. 
In this way, we shall extend to the ${\cal N}=0$ case the family of 
${\cal N}=1$
supersymmetry breaking terms first considered by Argyres and Douglas \cite{ad}.

We define the {\em spurion} variables $s_n$ as
follows
\be
s_1=-i \log \Lambda ~~~~~~~~~~~~~~~~~ 
s_n= -i \hat T_n ~, ~~~~~~ n=2, \dots, r=N-1.
\label{spurions}
\ee
Our independent coordinates in the prepotential are $\alpha^i, s_n$.
Using (\ref{completeness}) one can find explicit expressions for the dual
spurions:
\beqa
s_D^1 & = & {\beta\ov 2\pi } \biggl[ \H_2 +i \sum_{m \ge 2} m s_m \H_{m+1} 
- \sum_{m,n\ge 2} m s_m s_n \H_{m+1,n+1} \biggr] ~, \nonumber \\
s_D^n & = & {\beta\ov 2\pi  n} \biggl[ \H_{n+1} + i \sum_{m\geq 2} m s_m
\H_{m+1,n+1}\biggr] ~.
\label{dualspu}
\eeqa
Notice that, when the spurions $s_m$ are zero, we recover for the
variable $s_1$ the results of \cite{moresoft}.
Under the symplectic group ${\rm Sp} (2r, {\bf Z})$, 
the spurions are taken to be scalars, $s^\Gamma_m = s_m$. From the point
of view of the
Toda--Whitham hierarchy, this is natural in that the slow times parametrize
deformations of the curve, and should not be affected by duality
transformations (which
are transformations among symplectic basis of homology cycles of the curve). 
From the point of view of physics, this invariance is important because
their \vev
is an external unambiguous input.
We see from (\ref{dualspu}) that {\em the dual times are also invariant} under 
duality transformations.

To break ${\cal N} = 2$ supersymmetry down to ${\cal N} = 0$, as
anticipated above,
we promote the variables $s_n$ to ${\cal N} = 2$ vector superfields $S_n$, and
then freeze the scalar and auxiliary components to constant vacuum expectation
values.
We would like to restrict our framework to non-supersymmetric deformations of the
{\em original} pure $SU(N)$ super Yang--Mills theory.
Thus, for all $S_n, n \geq 2$, we only keep the top components $F_n$ as a
supersymmetry 
breaking parameter (by $SU(2)_R$ symmetry, we can always rotate the $D_n$
components 
away). In terms of ${\cal N}=1$ superfields we have,
\be
S \equiv S_1 = s_1 + \theta^2 F_1 ~~~~~~~~~~~~~~~~~ 
S_n = \theta^2 F_n ~,~~~~~~~ n\ge 2 ~,
\label{vevs}
\ee
where $s_1$ is related, as seen in (\ref{spurions}) to the dynamical scale
of the 
theory.  
The analysis of the soft breaking induced only by $S_1$ has been done
in Refs.\cite{soft,moresoft}.

As the prepotential has an analytic dependence on the spurion superfields,
the effective Lagrangian up to two derivatives and four fermion terms for
the ${\cal N} = 0$ theory is given by the exact
Seiberg--Witten solution once the spurion superfields are taken into account.
This gives the exact effective potential at leading order and the vacuum
structure can be determined. 
That is, all over the quantum moduli space, the effective action will be
\be
{\cal L}_{VM} = {1\ov 4\pi} \Im\left[
\int d^4 \theta \pder{F}{\Phi^I}\bar \Phi^I + {1\ov 2}\int d^2\theta
\ppder{F}{\Phi^I}{\Phi^J} W^I_\alpha W^{\alpha J}\right] ~,
\label{vecmu}
\ee
where the capital indices $I,J$ stand both for $i,j=1,...,N-1$ labelling abelian 
chiral ${\cal N}=2$ multiples $(\Phi^i, V^i)$, and for $m,n= 1,...,N-1$
that label
spurion multiplets $(S^m,V_s^m)$.
If we are near a submanifold of the moduli space of vacua where $n_H$
hypermultiplets become massless, the full Lagrangian also contains 
the hypermultiplet contribution involving pairs of chiral superfields
$H_a$, $\tilde H_a$, $a=1, \dots, n_H$,
\be
{\cal L}_{HM} = \sum_a \int d^4\theta \left( H^*_a e^{2 n^a_i V^i} H_a
+ \tilde H^*_a e^{-2 n^a_i V^i} \tilde H_a \right) + \sum_{a,i} \left( \int d^2 
\sqrt{2}\Phi^i n^a_i H_a \tilde H_a + \hbox{h.c.}\right) ~,
\label{lhm}
\ee
where the charge of the $a$-th hypermultiplet with respect to the $i$-th
$U(1)$ factor has been denoted by $n^a_i$ and, in the previous equation, a
particular choice of
duality frame, $a^i$, has been made. 
Namely the vector multiplets  $(\Phi^i, V^i)$, are such that near the singular
subvariety, the light BPS states in the previous lagrangian are weakly
coupled, and
perturbation theory is reliable. 
Of course, this amounts to an appropriate choice of the basis of homology
cycles $(A_i,B^j)$. 
Now, if the BPS states becoming massless are mutually local, we can always fix
a basis of cycles such that each $U(1)$ couples to one and only one
hypermultiplet. 
This means that $n^a_i=\delta^a_i$ or vanishes, the later case being
possible when 
$n_H<N-1$.

The full effective lagrangian will be the sum of (\ref{vecmu}) and (\ref{lhm}).
The effective potential can be computed explicitely resulting in
\beqa
V & = & B^{mn} F_m F^*_{n} + \sqrt{2}\,(n^a,b^m) \left(  F_m\tilde h_a h_a +
\bar F_m\bar h_a\bar{\tilde h}_a \right)
+ 2 (n^a,n^b)( h_a{\tilde h}_a \bar{h}_b \bar{\tilde h}_b) \nonumber\\
& + & {1\ov 2} (n^a,n^b)(|h_a|^2-|\tilde h_a|^2)(|h_b|^2-|\tilde h_b|^2)
+ 2 |n^a \cdotsh a|^2(|h_a|^2 + |\tilde h_a|^2) ~, 
\label{pfinal}
\eeqa
where $n^a \cdotsh a= \sum_i n^a_i a^i$, and $h_a$ ($\tilde h_a$) is the scalar 
component of $H_a$ ($\tilde H_a$). 
Also we have used the quantities
\be
(n^a,n^b) = n^a_i b^{-1\,ij} n^b_j ~~~~~~~~
(n^a,b^m) = n^a_i b^{-1\,ij} b_j{^m} ~~~~~~~~
B^{mn} = b_{a}{^m}{b}^{-1\,ab}b_b{^n} - b^{mn} ~,
\label{quantis}
\ee
where $b$ is given in terms of the generalized $2(N-1) \times 2(N-1)$ matrix of
couplings $\tau_{IJ}$,
\be
\tau_{ij}= {\partial^2 {\cal F}\over \partial \alpha^i \partial \alpha^j}
~~~~~~~~~~~~~~
\tau^n{_i}= {\partial^2 {\cal F}\over \partial \alpha^i  \partial  s_n}
~~~~~~~~~~~~~~
\tau^{mn}= {\partial^2 {\cal F}\over \partial s_m \partial s_n} ~,
\label{coupling}
\ee
as
\be
b_{IJ} = {1 \over 4 \pi} {\rm Im}\,\tau_{IJ} ~,
\label{imcop}
\ee
and it can be computed all over the moduli space from
Eqs.(\ref{mixed})--(\ref{secorder}). This is precisely the point where the
Seiberg--Witten solution enters the calculations.

To obtain the values of the condensates, we first minimize $V$ with
respect to $h_a$,
$\tilde h_a$, resulting in $|h_a| = |\tilde h_a|$. 
It is convenient to fix the gauge in the $U(1)^{N-1}$ factors in such a way that
\be
h_a = \rho_a  ~~~~~~~~~~~~~~~~~ \tilde h_a = \rho_a e^{i\beta_a} ~.
\label{gaugefix}
\ee
If the charge vectors $n^a$ are linearly independent, the non-trivial condensates
satisfy the equation
\be
|n^a \cdotsh a|^2 + \sum_b(n^a,n^b)\rho_b^2
e^{i(\beta_b-\beta_a)}
 + {1\ov \sqrt{2}}(n^a,b^m) F_m  e^{-i \beta_a } = 0 ~,
\label{condensados}
\ee
and the effective potential takes the value
\be
V= B^{mn}F_m F^*_n - 2
\sum_{ab}(n^a,n^b)\rho_a^2\rho_b^2
\cos(\beta_a- \beta_b) ~.
\label{potefftres}
\ee
We will consider in what follows, as an explicit example, the case of $SU(3)$.

%%%%%%%%%%%%%
%%%%%%%%%%%%%
\section{Analysis of $SU(3)$}
%%%%%%%%%%%%%
%%%%%%%%%%%%%

The ${\cal N}=2$ super Yang-Mills theory with gauge group $SU(3)$ has been
analyzed in detail in Refs.\cite{klemm,ad}.
There are two sets of distinguished singularities in the moduli space of
this theory: \\
i. The three ${\bf Z}_2$ vacua, known as ${\cal N}=1$ or maximal points
\cite{ds}, located at $u^3=27\Lambda^6/4$, $v=0$, that give rise to the
${\cal N}=1$ vacua when the
theory is perturbed with a mass term of the form ${\rm Tr} \Phi^2$ (we
denote $u_2=u$, $u_3=v$). \\ 
ii. The two ${\bf Z}_3$ vacua, known as Argyres--Douglas (AD) points \cite{ad},
located at $u=0$ and $v=\pm 2\Lambda^3$, where three mutually nonlocal BPS states
become simultaneously massless. The low-energy theory there is an ${\cal N}=2$
superconformal theory. \\ We will briefly describe the situation near both
kind of singularities. We set $\Lambda^{6}=4$ for convenience.

\subsection*{The ${\bf Z}_2$ vacua}

In this subsection we study the soft breaking of the theory near the
${\cal N}=1$
points where two magnetic monopoles become simultaneously massless.
To evaluate the second derivatives of the prepotential, we need the values of the
periods of the hyperelliptic curve and the structure of the gauge couplings.
We will focus on the ${\cal N}=1$ point ($u=3$,$v=0$), whereas the values of the
quantities at the other two points can be obtained by using the ${\bf
Z}_3$ unbroken symmetry. 
The derivatives of the Casimir operators with respect to the dual variables are
given by \cite{emm}
\be
{\partial u  \over \partial a_{D\,j}}=-2i \sin { \pi j \over N}~~~~~~~~~~~~~~~
{\partial v  \over \partial a_{D\,j}}=-2i \sin { 2\pi j \over N} ~.
\label{derivs}
\ee
The gauge couplings near the ${\cal N}=1$ point have the structure
\be
\tau^D_{ij}= {1 \over 2\pi i } \log \left( {a_{D\,i}\over 2\sqrt{3}}  \right)
\delta_{ij} + (1-\delta_{ij})\tau_{ij}^{\rm off} + {\cal O}(a_{D\,i}) ~,
\label{coupl}
\ee
where $\tau_{ij}^{\rm off}$, $i \not=j$ are the off-diagonal entries of the 
coupling constant at the ${\cal N}=1$ point, which will be discussed in
some detail 
in the next section.  
For $SU(3)$, $\tau_{12}={i \over \pi} \log 2$ \cite{klemm}.
To compute the Theta function, we have to take into account the change of
the {\em
electric} characteristic under the symplectic transformation to the
magnetic variables,
$a_{D\,i}$. 
Indeed, the transformation law for the Theta function is given by \cite{rauch}
\be
\Theta [\alpha^{\Gamma}, \beta^{\Gamma}] (\tau^{\Gamma}|
\xi^{\Gamma})= {\rm e}^{i \phi} ({\rm det} (C\tau + D))^{1/2}
\exp \bigl[ \pi i \xi^t (C\tau + D)^{-1} C \xi \bigr] 
~\Theta [\alpha , \beta](\tau|\xi) ~,
\label{thetatrans}
\ee
where $\phi$ is a $\xi$-independent phase and
\be
\alpha^{\Gamma}= D\alpha-C\beta +{1 \over 2} {\rm diag}(CD^t) ~~~~~~~~~~~~
\beta^{\Gamma}= -B\alpha + A \beta+ {1 \over 2} {\rm diag}(AB^t) ~.
\label{chartrans}
\ee
Thus, the {\em magnetic} or {\em dual} characteristic, D, is
\be
\vec \alpha= (1/2, 1/2) ~~~~~~~~~~~~~~~~~~~~~~~~~~~~~~ \vec \beta=(0, 0) ~.
\label{dualchar}
\ee
From the leading behaviour of the Theta function with dual characteristic, we can
compute its derivative at the ${\cal N}=1$ point of $SU(3)$,
\be
\left.{1\ov i\pi} \d_{\tau_{ij}} \log\Theta_D(0|\tau^D)\right\vert_{a_{D\,i}=0}
= {1\ov 4}\delta_{ij} - {1\ov 12}(1-\delta_{ij}) ~.
\label{nonetheta}
\ee
Using (\ref{derivs}) and (\ref{nonetheta}), it is easy to check the relation
(\ref{qucas}) for the $\Lambda$ derivatives of the Casimir operators.

At the ${\cal N}=1$ point, there is a symplectic basis for the
hyperelliptic curve,
such that the magnetic charge vectors are given by $n^a_j=\delta^a_j$,
$a_{D\,i}=0$,
and from eq.(\ref{condensados}), the condensates  are given by
\be
\rho^2_1 = \sqrt{3\ov 2} \,{3\ov 4\pi^2} \,|F_1 + \med F_2| ~~~~~~~~~~~~~~
\rho^2_2 = \sqrt{3\ov 2} \, {3\ov 4\pi^2} \,|F_1 - \med F_2| ~. 
\label{conde}
\ee
We see that the soft breaking induced by the quadratic and cubic Casimirs
gives rise 
to monopole condensation in both $U(1)$ factors, although the condensates are
bigger for the soft breaking coming from $u$ (for equal values of the
supersymmetry
breaking parameters $F_1$, $F_2$).  
In the same way, the vacuum energy associated to these condensates is
\be
V_{\rm eff} = - b^{mn} F_m F^*_n = 
-{9\over 4\pi ^2} \left( |F_1|^2 + \med |F_2|^2\right)~.
\label{vacen}
\ee
As expected, the soft breaking associated to $u$ gives lower energy to
this vacuum.

\subsection*{The ${\bf Z}_3$ vacua}

Next we explore the behaviour near the Argyres--Douglas point at ($u=0$, $v=4$).
It is convenient to use the parameters $\rho$ and $\epsilon$ already
introduced in Ref.\cite{ad},
\be
u= 3 \epsilon^2 \rho ~~~~~~~~~~~~~~~~~~~~~~~~~~~~ v-4 = 2 \epsilon^3 ~.
\label{rhoeps}
\ee
The three submanifolds $\rho^3=1$ correspond to three massless BPS states which
after an appropriate symplectic transformation can be seen to be charged
with respect
to only one of the $U(1)$ factors, with variables denoted by $a^1, a_{D\,1}$.
They can be seen to be an electron, a dyon, and a monopole. 
These submanifolds come together at the AD point, where a nontrivial
superconformal
field theory is argued to exist \cite{ad}.
To leading order, the hyperelliptic curve splits at the AD point into a
small torus
(corresponding to two mutually nonlocal periods $a^1, a_{D\,1}$ which go
to zero) and a
big torus with periods $a^2, a_{D\,2}\sim\Lambda$.
The small torus is given by the elliptic curve
\be
w^2=z^3-3 \rho z -2 ~,
\label{rhocurve}
\ee
and the meromorphic Seiberg--Witten differential degenerates on it to
\be
\lambda_{SW} = {1 \over 2\sqrt{2}\pi} \epsilon^{5/2} w dz ~.
\label{swdif}
\ee
The matrix of couplings near the AD point, at leading order, reads 
\cite{ad,moore2,ky}
\be
\tau_{11} = \tau(\rho) + {\cal O}(\epsilon) ~~~~~~~~~~
\tau_{12} = - {i \over c} {\epsilon^{1/2}\over
\omega_{\rho}}+ {\cal O}(\epsilon^{3/2}) ~~~~~~~~~~
\tau_{22} = \omega + {\cal O}(\epsilon) ~, 
\label{couplings}
\ee
where $\omega_\rho$ is the period of the small torus (with ${\rm
Im}(\omega_{\rho\,D}/\omega_{\rho})>0$), 
$c$ is a nonzero constant and $\omega= {\rm e}^{\pi i/3}$.
For the dual variables we have similar expressions with $\omega_{\rho\,D}$
and $c_D$. 

To analyze the Theta function in these variables, we need the appropriate
characteristic which, using (\ref{chartrans}) and the results in \cite{moore2},
is
\be
\vec \alpha= \vec\beta= (1/2,1/2) ~.
\label{adchar}
\ee
We can already obtain the behaviour of the Theta function as an expansion in
$\epsilon$:
\be
\Theta(0|\tau)  = -{1 \over 2\pi c}  {\epsilon^{1/2}\over
\omega_{\rho}} \vartheta_1'(0|\tau(\rho)) \vartheta_1'(0|\omega) + {\cal
O}(\epsilon^{3/2}) ~,
\label{theta}
\ee
where $\vartheta_1(\xi|\tau)$ is the Jacobi theta function with characteristic
$[1/2,1/2]$. 
Now, using that
\be
{\vartheta_1'''(0|\tau)\over \vartheta_1'(0|\tau) } =-\pi^2 E_2(\tau) ~,
\label{ident}
\ee
we find the leading contribution to the derivative of the Theta function
\be
{1 \over i\pi} \partial_{\tau_{ij}}\log \Theta = \bemat{cc} {1\over 4} E_2
(\tau(\rho)) & {c \over 2\pi} \epsilon^{-1/2} \omega_{\rho} \\
{c \over 2\pi} \epsilon^{-1/2} \omega_{\rho} & {1 \over 4} E_2(\omega) \enmat ~.
\label{thetamat}
\ee
Again, it can be checked that the relation (\ref{qucas}) for $v$ holds (for $u$, 
it is necessary to know the explicit values of the constants).

The analysis of the condensates near the AD point is difficult because one has
to take into account mutually nonlocal degrees of freedom, and there is not a 
Lagrangian description of this theory. 
In fact, one expects that, in the softly broken theory, a cusp singularity
will appear
in the effective potential near the AD point, as it happens in ${\cal N}=2$
QCD with gauge group $SU(2)$ and one massive flavour \cite{luisIyII}. 
But we can analyze the monopole condensates along the divisors $\rho^3=1$
and their
evolution as we approach the AD point.
Near each of the submanifolds $\rho^3=1$ there is a massless BPS state, and we
expect it to condense after breaking supersymmetry down to ${\cal N}=0$. 
These condensates correspond to mutually nonlocal states but we can
assume, following
the discussion in Refs.\cite{moresoft,luisIyII}, that these states do not
interact,
the condensates being given by the equation \cite{emm}
\be
\rho^2_k=-{1 \over (b^{-1})_{11} }|a_k|^2 - {{\rm e}^{-i\beta_k} \over {\sqrt
2} (b^{-1})_{11} } \sum_{n=1,2}F_n
(b^{-1})_{kj} b^n_{~j} ~,
\label{rhoconds}
\ee
where $k=1,2,3$ and $a_k$ are the appropriate local coordinates for each of the
massless states ({\it i.e.} $a_k=a^1$, $a_{D\,1}$, $a^1-a_{D\,1}$). 
The quantities $(b^{-1})_{ij}$, $b^n{_j}$ should be also computed in the
duality frame dictated by the $a_k$. 
This approximation should be good far enough from the AD point. 
These condensates give only a magnetic Higgs mechanism in one of the
$U(1)$ factors, 
and correspond to the half-Higgsed vacua of \cite{ad}. 
Notice that one should perform a careful numerical study of the equations for the
condensates and for the effective potential to know if these partial
condensates give
the true vacua of the ${\cal N}=0$ theory. 
As we approach the AD point, $\epsilon \rightarrow 0$, we see that the
parameters for
condensation go to zero for both the quadratic and the cubic Casimirs:
\be
{\partial u\over \partial a^1}, \,\ {\partial v\over \partial a^1} \sim {\cal
O}(\epsilon^{1/2}),
\label{gozero}
\ee
and the mass gap associated to the condensates vanishes at the AD point, like
in the ${\cal N}=1$ breaking considered in Ref.\cite{ad}.

%%%%%%%%%%%%%
%%%%%%%%%%%%%
\section{Strong Coupling Expansion near the Maximal Points}
%%%%%%%%%%%%%
%%%%%%%%%%%%%

Let us end by applying the Seiberg--Witten--Whitham equations in the
strong coupling regime of ${\cal N}=2$ super Yang-Mills theory near its
maximal singularities. 
The case of $SU(2)$ is special in that, being its Cartan subalgebra
one--dimensional, the whole strong coupling expansion of the prepotential
can be recursively computed {\em without} an explicit knowledge of the
actual solution $(a(u),a_D(u))$ \cite{edemas}, much in the same way than
the previously derived instanton corrections.
For generic $SU(N)$, however, the SWW equations do not give a closed procedure
to obtain the strong coupling expansion of the effective prepotential.
Aside from some technical difficulties, the main problem is that, in spite of the
fact that a grading in $a_{D\,i}$ and $\Lambda$ still exists, higher terms of the
expansion appear in the equations corresponding to lowest powers of the
dual variables spoiling recursivity (see Ref.\cite{edemas} for details).
The SWW equations do not seem to be instrumental to study the full strong
coupling expansion of $SU(N)$ ${\cal N}=2$ super Yang--Mills theory.

Other methods have been derived in the literature to tackle the problem of
computing the higher threshold corrections to the effective prepotential.
For example, in Ref.\cite{dHP} this has been accomplished by parametrizing the
neigborhood of the maximal singularities with a family of deformations of the
corresponding auxiliary (singular) Riemann manifold.
However, this formalism is not sensitive to quadratic terms in the prepotential. 
Thus, in particular, it does not give an answer for the couplings between
different magnetic $U(1)$ factors at the maximal singularities of the
moduli space, $\tau_{ij}^{\rm off}$, introduced in the previous section.
The existence and importance of such terms has been first pointed out in
Ref.\cite{ds} by using a scaling trajectory that smoothly connects the
maximal singularities with the semiclassical region.
These terms are also important ingredients in the expression of the
Donaldson--Witten functional for gauge group $SU(N)$ \cite{moore2}.
To our knowledge, a closed formula for these off-diagonal couplings has not been
obtained so far, except for the gauge group $SU(3)$ \cite{klemm}.
Let us then consider the uses of the SWW equations in the solution of this
problem.

We have seen before that physical quantities in the neighborhood of any maximal
singularity can be translated to a patch in the vicinity of any other by the
action of the unbroken discrete subgroup ${\bf Z}_N$. 
We will consider in what follows the point where $u_2$ is real and positive.
The strong coupling expansion of the prepotential at such singular point can 
be written in terms of appropriate $a_{D\,i}$ variables as\footnote{We follow 
here the conventions of Ref.\cite{ds} to fix the first three terms of the 
expansion.}
\[ \F = \frac{N^2}{2\pi i}\Lambda^2 + 
\frac{2N\Lambda}{\pi}\sum_{k=1}^{N-1} a_{D\,k}\,\sin\hat\theta_k + {1\ov 4\pi 
i}\sum_{k=1}^{N-1}a_{D\,k}^2 \log \frac{a_{D\,k}}{\tilde\Lambda_k} + \frac{1}{2}
\sum_{k\neq l=1}^{N-1} \tau_{kl}^{\rm off}\,a_{D\,k} a_{D\,l} 
+ {1\ov 2 \pi i}\sum_{s=1}^\infty \F_s(a_D)\Lambda^{-s} ~, \]
where $\hat\theta_k = \pi k/N$ and the logarithmic term, coming from the one-loop
diagram that involves the light monopole, has the appropriate sign and
factor making
manifest that the theory is non-asymptotically free and that there is a monopole
hypermultiplet weakly coupled to each dual photon for $a_{D\,i}\to 0$.
The remaining power series expansion comes from the integration of
infinitely many massive BPS states: $\F_s(a_D)$ are polynomials of degree
$s+2$ in dual variables and 
$\tilde\Lambda_k = e^{3/2}\Lambda\sin\hat\theta_k$.

We have seen that the SWW formalism allow us to relate the strong coupling
expansion
of homogeneous combinations of higher Casimir operators $\H_m$ with that of the
prepotential through Eq.(\ref{qucas}).
Let us first remark that this equation is also valid for the higher
Casimirs $h_n$ themselves \cite{ITEP}:
they, as well as their particular combinations encoded in $\H_n$, are
homogeneous
functions of $a_D$ and $\Lambda$ of degree $n$.
Thus, at the ${\cal N}=1$ singularities, the LHS of Eq.(\ref{qucas}) is simply
\be
{\partial h_n  \over \partial\log\Lambda} = n h_n =
\sum_{k=1}^{N} (2\cos\theta_k)^n ~,
\label{LHS}
\ee
where we used the fact that the eigenvalues of $\phi$ are $\phi_i = 
2\cos\theta_i$ with $\theta_i = (i-1/2)\pi/N$ \cite{ds}.
The derivative of the Casimir operators with respect to the dual variables 
can be computed at the same point of the moduli space, by using the explicit 
representation of the curve in terms of the Chebyshev polynomials \cite{emm},
resulting in
\be
{\partial h_n  \over \partial a_{D\,j}}= -2i \sum_{l=0}^{[n/2-1]}
{n-1 \choose l} \sin (n-2l-1)\hat\theta_j ~,
\label{derHsun}
\ee
whereas, from the expansion of the effective prepotential, the leading
couplings at
the maximal singularity are given by
\be
\tau^D_{ij }= {1 \over 2\pi i } \log \left( {a_{D\,i}\over \Lambda_i}  \right)
\delta_{ij} + \tau_{ij}^{\rm off} ~.
\label{latau}
\ee
The derivative of the Theta function $\Theta_D$ with respect to the period 
matrix has the following expression when evaluated at the ${\cal N}=1$ 
singularity
\be
{1\ov i\pi}\partial_{\tau^D_{ij}}\log \Theta_D(0,\tau_D) = \frac{1}{4} 
\left(\sum_{\xi^k=\pm 1}\exp{(i\frac{\pi}{4}\xi^l\tau_{lm}^{\rm 
off}\xi^m)}\right)^{-1} 
\sum_{\xi^k=\pm 1}\xi^i\xi^j\exp{(i\frac{\pi}{4}\xi^l\tau_{lm}^{\rm 
off}\xi^m)} ~.
\label{dlogsun}
\ee
Now, we can insert the results (\ref{LHS})--(\ref{dlogsun}) in the SWW equations
(\ref{qucas}) obtaining
\beqa
\frac{1}{2N}\sum_{k=1}^{N} (2\cos\theta_k)^n & = & \sum_{l=0}^{[n/2-1]}
{n-1 \choose l} \sin\hat\theta_i \sin (n-2l-1)\hat\theta_j
\left(\sum_{\xi^k=\pm 1}\exp{(i\frac{\pi}{4}\xi^l\tau_{lm}^{\rm 
off}\xi^m)}\right)^{-1} \nonumber \\ & & ~~~~~~~~ \times 
\sum_{\xi^k=\pm 1}\xi^i\xi^j\exp{(i\frac{\pi}{4}\xi^l\tau_{lm}^{\rm 
off}\xi^m)} ~.
\label{final}
\eeqa
We have $N-1$ equations and $(N-1)(N-2)/2$ unknowns (the components of the 
symmetric matrix $\tau_{ij}^{\rm off}$). 
Thus, Eq.(\ref{final}) has predictive power in its own only for $SU(3)$
and $SU(4)$. 
Indeed, we obtain for these two cases the following values:
\be
SU(3): ~~~ \tau_{12}^{\rm off} = i/\pi \log{2} ~~~~~~~~~~~~
SU(4): ~~~ \left\{ \begin{array}{l} \tau_{12}^{\rm off} = 
\tau_{23}^{\rm off} = - i/\pi \log(\sqrt{2}-1) \\
\tau_{13}^{\rm off} = i/\pi \log\sqrt{2} \end{array} \right. ~.
\label{su4}
\ee
Notice that our result for $SU(3)$ coincides with that of Ref.\cite{klemm}
while the
ones for $SU(4)$ have not been found previously.
For higher $SU(N)$, further ingredients would be necessary in order to obtain the
off-diagonal couplings at the ${\cal N}=1$ singularity.
Instead, we can think of Eq.(\ref{final}) as a {\em new} constraint that
$\tau_{mn}^{\rm off}$ must obey.
In fact, inspired by the findings in the last section of Ref.\cite{ds}, we
{\em propose} the following ansatz for $\tau_{mn}^{\rm off}$:
\be
\tau_{mn}^{\rm off} = \frac{2i}{N^2\pi} \sum_{k=1}^{N-1} \sin k\hat\theta_m 
\sin k\hat\theta_n \sum_{i,j=1}^N \tau_{ij}^{(0)} \cos k\theta_i 
\cos k\theta_j ~,
\label{result}
\ee
with $\tau_{ij}^{(0)}$ being given by
\be
\tau_{ij}^{(0)} = \delta_{ij} \sum_{k\neq{i}}\log(2\cos\theta_i-2\cos\theta_k)^2
- (1-\delta_{ij}) \log(2\cos\theta_i-2\cos\theta_j)^2 ~.
\label{logs}
\ee
There is no equivalent expression available in the literature to compare with. 
Nevertheless, we can use precisely the SWW equations (\ref{final}) in order to
make a non-trivial check of our ansatz for the off-diagonal couplings
(\ref{result})--(\ref{logs}). 
We have done it numerically up to SU(11) with remarkable sucess \cite{edemas}. 
There is a second check that we can do using results that do not rely on  Whitham
equations at all. Douglas and Shenker showed that the matrix
$\tau^D_{mn}$ at any point of the scaling trajectory, diagonalizes in the
basis $\{\sin k\hat\theta_n\}$ with  certain particular eigenvalues (see
Eqs.(5.9)--(5.12) of Ref.\cite{ds}). The couplings
(\ref{result})--(\ref{logs}) satisfy this restrictive
condition  in the limit of the scaling trajectory ending at the maximal
singularity. As
long as our solution (\ref{result})--(\ref{logs}) matches two very stringent and
independent conditions, we believe that it provides a faithful answer for
$\tau_{mn}^{\rm off}$ as well as a highly non-trivial test of the 
Seiberg--Witten--Whitham formalism. 

%%%%%%%%%%%%%
%%%%%%%%%%%%%
\section*{Acknowledgements}
%%%%%%%%%%%%%
%%%%%%%%%%%%%

We are indebted to Marcos Mari\~no for many insightful discussions.
J.D.E. would like to thank the organizers of the Meeting {\em Trends in
Theoretical Physics II}, for giving him the opportunity to present these
results. 
The work of J.D.E. is supported by a fellowship of the Ministry of 
Education and Culture of Spain. The work of J.M. was partially supported
by DGCIYT under contract PB96-0960.

%---------------- Bibliografia-------------------

\end{document}